# InP optical amplifiers with Euler U-bend waveguide geometry for low-loss flip-chip hybrid integration

HEIDI TUORILA,[1,*] JUKKA VIHERIÄLÄ,[1] LEE JAE-WUNG,[2] MIKKO HARJANNE,[2] MATTEO CHERCHI,[2,3] TIMO AALTO,[2] AND MIRCEA GUINA[1]

[1] *Optoelectronics Research Centre, Physics Unit, Tampere University, 33101 Tampere, Finland*
[2] *VTT Technical Research Centre of Finland, Espoo 02044, Finland*
[3] *Current affiliation Xanadu Quantum Technologies*

**Abstract:** We report on the development of InP-based semiconductor amplifiers with a U-bend waveguide geometry having the input and output ports on one facet only. This waveguide geometry simplifies the chip alignment during the hybrid integration on silicon photonics platforms ultimately reducing the coupling losses, improving the integration yield, and minimizing the length of the optoelectronic chip. To achieve low loss U-bends with small footprint, we utilize the Euler bend geometry previously demonstrated on silicon and GaAs platforms. We analyze the gain properties of the devices by operating them as laser diodes at room temperature. Low loss U-bend performance with a 0.56 dB for a 50 µm effective bending radius bend in a single-mode strip InP waveguide is demonstrated. The interface between bend and straight waveguides was studied by comparing deep etched waveguides to a combination of shallow straight waveguides and deep etched bends. The effects of this interface on the device losses, electric properties and spectrum are reported. The implications related to having a bend section on the carrier injection and gain are discussed. Finally, results on the integration trials on silicon-on-insulator platform are presented.

## 1   Introduction

Photonic integrated circuits (PIC) are gradually maturing and reaching a wider range of applications beyond the established datacom use cases [1]. While silicon photonics (SiPh) platforms form the current backbone of the PIC development, silicon's lack of gain and light emitting capabilities [2] has triggered intensive development for integrating III-V gain heterostructures supporting these functions.

Hybrid integration of SiPh and III-V optoelectronic components exploiting flip-chip bonding provides a wavelength versatile and flexible integration method required to tackle the increasing range of application-related functions. While flip-chip integration is seen to be less cost-effective and to offer a lower integration density solution compared to the heterogenous [3] or monolithic integration [4], it enables fabrication of custom PIC functionalities exploiting optimized SiPh and III-V components. The challenge with the hybrid integration are the strict requirements for accurate device placement. For travelling wave (TW) optoelectronic devices, such as semiconductor optical amplifiers (SOA) this results in a need for a precise control of the chip dimensions to match the alignment with the predefined SiPh butt-coupling waveguide interfaces [5,6]. Generally, to achieve a high coupling ratio, the longitudinal gap between the III-V and SiPh interfaces must be below one micrometer. However, we should note that using typical dicing equipment yields chip lengths with variation as high as 5-10 µm [6]. This ultimately results in high coupling losses or even in impossibility to mount the chip within the SiPh predefined flip-chip bonding areas when this type of inaccuracy leads to a longer device length. To this end, advanced processing techniques, in which the device length is defined by lithography, have been developed [5,6] yet the need of an accurate alignment at both waveguide ends remains. The solution we adopt in this paper is the use of a U-bend waveguide design to bring the I/O ports on the same facet, which allows a simplified alignment at one end of the waveguides with precise alignment and control, limited only by the precision of the flip-chip bonding tools. The concept is illustrated in Figure 1.

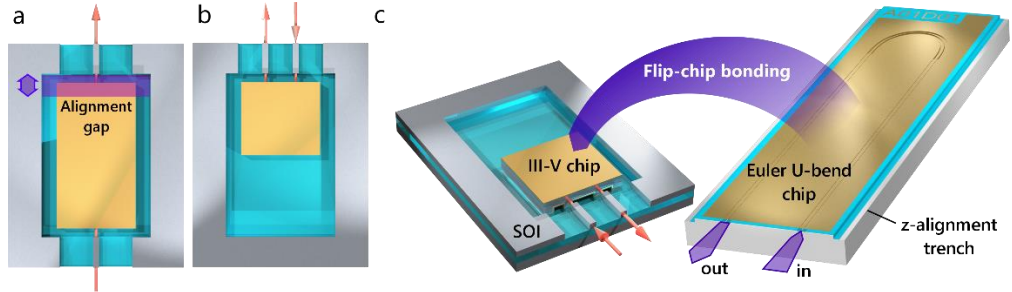

Figure 1. a) TW chip with device length variance induced alignment gap variance vs. b) U-bend chip with precise control of the alignment gap. c) Euler U-bend travelling wave gain device flip-chip integration concept.

The U-bend active waveguide concept has been deployed for hybrid integration of GaAs/SOI waveguides [7] and also reported for the fabrication of a circular InP U-bend SOA with a 75 µm bending radius [8]. However, the use of the U-bend design is not without its own problems. Usually, circular bends require a large bending radius and narrow waveguides to maintain low-loss, single-mode operation [9–12]. For this reason we have adopted an Euler bend design, which was initially introduced to reduce the bending loss on silicon waveguides on a 3 µm core silicon-on-insulator (SOI) platform with the intention to fabricate low-loss, single-mode bends [11]. This approach allows for a µm-scale technology with integration density equaling the 220 nm SOI platform narrow waveguides [11] without increasing the losses. We have exploited this design in the fabrication of low-loss Euler U-bend TW gain structures based on GaAs/GaInNAs heterostructures [7]. In this work, the Euler U-bend concept is expanded to InP-material system and semiconductor amplifiers operating at telecom wavelengths. Low-loss U-bend SOAs with an effective bending radius of 50 µm flip-chip integrated on an SOI platform are demonstrated.

In our previous work on GaAs Euler U-bends it was noted that similarly to Si waveguides, it is necessary to utilize deep etched strip waveguides in the bend areas to provide high mode confinement for low-loss operation. However, the fabrication of deeply etched waveguides is known to exhibit surface related detrimental issues, such as increased leakage currents and non-radiative carrier losses that reduce the device performance [13–16]. Thus, in theory it would be preferrable to use the strip design only in the bend sections. Because of this, a design with a deep etched U-bend and shallow etched straight sections was utilized here to ensure maximum net gain. Subsequently, in this work one of the goals was to further test and analyze the effect of the bend and the deep/shallow etch interface on the gain device performance.

## 2 Device design and fabrication

Both SOAs and laser diodes (LDs) were fabricated for the study reported. Although the U-bend LDs are not relevant for the end applications, they provide an insight to the U-bend performance. The test series is illustrated in Figure 2. The series consists of different cavity length straight and U-bend devices with both deep etched strip and shallow etched rib waveguide profiles. By comparing the straight and U-bend device LD operation the effect of the etch depth and the step interface that is formed between the strip and rib sections (shown in Figure 3) can be assessed.

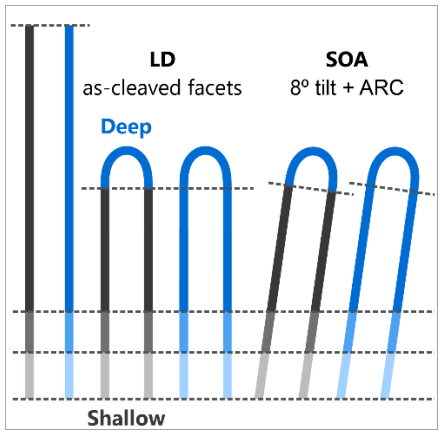

Figure 2 Device types used in the work.

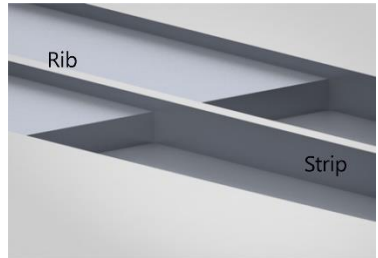

Figure 3 Rendered illustration of the he rib-strip interface in the U-bend devices.

The active structure was a commercially grown InP wafer and included 6 quantum wells surrounded by $In_{0.53}Al_xGa_{0.47-x}As$ waveguides and p-InP and n-$In_{0.52}Al_{0.46}As$ claddings. The deep etched sections were realized using a double-lithography self-aligning approach. The shallow sections were first dry etched in an inductively coupled plasma reactive ion etch system (ICP-RIE) using a dielectric hard mask, followed by a second lithography and etch step with a photoresist mask on top of the original hard mask so that the photoresist mask was opened in the areas where the deep etch was required. Besides this, an additional feature for the flip-chip bonding z-alignment control was added. To control the z-positioning of the III-V chip, a trench was dry etched at the device sides with corresponding z-stoppers formed on the SOI side. The etch depth was calculated so that the optical axis of the InP chip matched the SOI optical axis.

The rest of the fabrication was typical, including the deposition of the electrical contacts, flip-chip interface, and facet formation. A $SiO_2$ insulator layer deposited by plasma enhanced vapor deposition (PECVD) was used to limit the current injection region. A p-contact metal layer of Ti/Pt/Au was e-beam evaporated and patterned using a lift-off process. For the devices intended for flip-chip bonding a second 2 µm thick Au layer was electroplated. Figure 4 (a) and (b) show examples of both non-electroplated and electroplated devices, respectively. After this, the sample was thinned to 110 µm through lapping and polishing before the n-contact metal stack Ni/Au/Ge/Au was e-beam evaporated on the polished surface and alloyed using rapid thermal annealing. Finally, the devices were diced using a scribe and break system. The SOA devices were cleaved with an 8° tilt angle and were antireflection coated to prevent back reflections. The test devices were bonded with silver epoxy on sub-mounts and wire bonded for the characterization. The SOA devices intended for integration with the SOI platform were sent to a commercial fab for flip-chip bonding where they were soldered using the thick, electroplated Au layer on the III-V chips and Zn on the SOI, which together form an AuZn alloy.

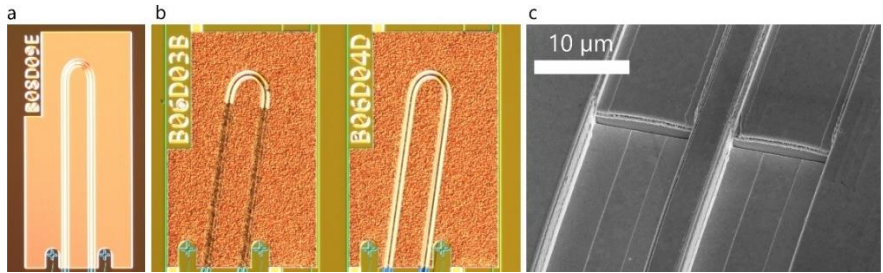

Figure 4 a) U-bend strip LD test devices without electroplating. b) Electroplated U-bend SOA devices for flip-chip bonding: strip waveguide and rib-strip waveguide. c) SEM image of the fabricated rib-strip interface.

## 3    Device characterization and analysis

### 3.1    Light-current-voltage characteristics

The light-current-voltage (LIV) characteristics were measured at 20°C in continuous wave (CW) mode. Examples of the LI and IV data are displayed in Figure 5 for each waveguide configuration, i.e. straight and U-bend LDs with rib and strip geometries. For the comparison with the straight devices, the output power for the U-bend devices is halved to account for having the both ports on the same facet. The measured devices had several different cavity lengths. The inverse of differential quantum efficiency (DQE) was plotted as a function of the cavity length. Figure 6 displays the plots and linear fits. Following the typical analysis [17], Table 1 summarizes the significant parameters that can be retrieved from these measurements.

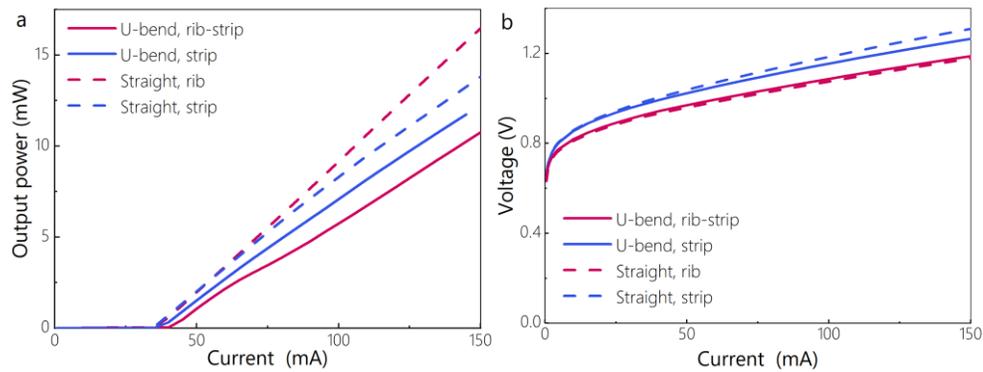

Figure 5 a) Examples of LI-curves for 1166 µm and 1151 µm straight and U-bend LD measurements. b) The VI-curves.

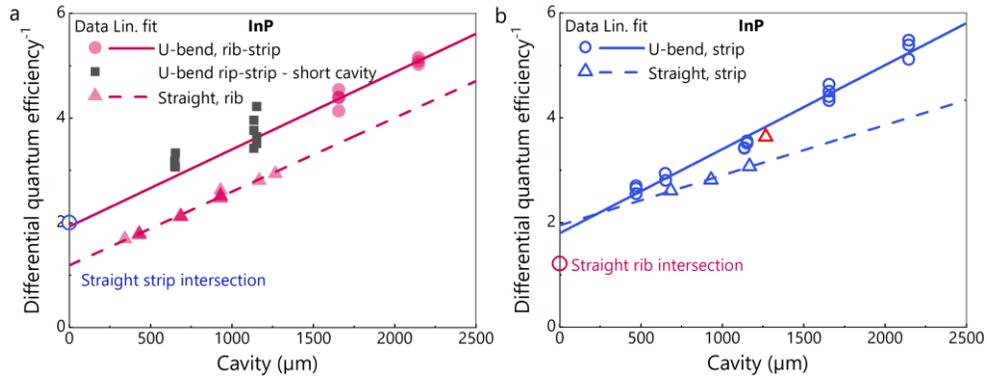

Figure 6 The data plots and linear fits for differential quantum efficiencies as a function of the cavity length for a) straight rib and U-bend rib-strip LD devices and b) straight and U-bend strip devices. The red, hollow triangle marks a poor-quality device ignore in the linear fit.

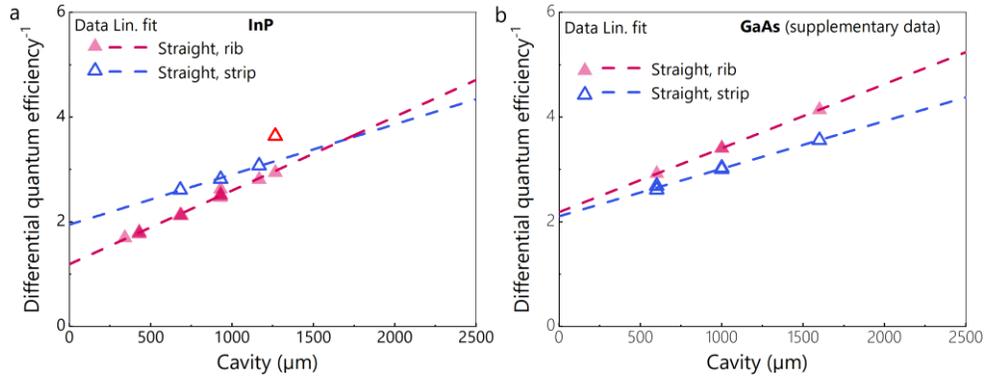

Figure 7 Comparison of 1/DQE fits between straight rib and strip devices on a) InP and b) GaAs material systems. The red, hollow triangle marks a poor-quality device ignored in the linear fit.

**Table 1 Summary of calculated LD device characteristics.**

|  | Cavity length (μm) | $J_{th}$ (A/cm$^2$) | DQE | IQE | $\langle \alpha_i \rangle$ (1/cm) | $\langle g_{th} \rangle$ (1/cm) |
|---|---|---|---|---|---|---|
| **Straight shallow** | 1166 | 1035 | 0.36 | 0.84 | 15.19 | 26.19 |
| **U-bend shallow** | 1151 | 1198 | 0.27 | - | - | - |
| **Straight deep** | 1166 | 1035 | 0.33 | 0.52 | 6.32 | 17.32 |
| **U-bend deep** | 1151 | 1198 | 0.28 | 0.56 | 11.41 | 22.56 |

## 3.2  Performance analysis

Analyzing the LI data shown in Figure 5 (a) we can see that the rib-strip U-bend devices have unexpectedly the lowest slope and exhibit a kink, while the initial hypothesis was that the strip U-bend should have the poorest performance. The results reveal that the rib-strip device type with the straight interface (Figure 4 (c)) seems to be affected the most by the process. This indicates that the interface causes significant losses for the device whereas when the deep plasma etch is used, while introducing bend losses, it is less harmful for the performance.

Figure 5 (b) displays the measured IV-characteristics and from them we can see that the deep etched devices suffer from a higher contact resistance. This can be attributed to the fabrication process related problem where the photoresist mask failed prematurely in the areas of deep etched RWGs exposing the waveguide top to a plasma etch unnecessarily.

The U-bend related bend losses were estimated using the approach introduced in [7] for GaAs Euler U-bend devices that is based on the observed difference in the slopes of the linear fit, see Figure 6 (b). From the comparison between the straight and U-bend strip devices an estimate of 5.7 1/cm or 0.56 dB for the 228 µm long U-bend bend loss was reached. As a specific point of interest, it can be noted that the Figure 6 (a) deviates from the behavior observed for the GaAs devices. The rib-strip U-bend plot appears to suffer from a non-linear behavior with the shorter cavity lengths where the rib-strip interface related effects dominate and are thus ignored in the linear fit. The reason for this is assumed to be the combination of non-radiative recombination and increased leakage currents having a stronger effect on shorter cavities, further enhanced by the rib-strip interface that forms a short sub-cavity [17–19].

The straight rib devices on the other hand exhibit much lower value for the y-axis interception point which corresponds to the inverse of the internal quantum efficiency (IQE) that is the measure of injected charge carriers contributing to the stimulated emission. By comparing the replotted data for the straight rib and strip devices to previously unpublished data available for a GaAs device series with straight rib and strip waveguides in Figure 7 (b), we see that for GaAs devices the plots converge to a single point at the y-axis. This indicates that the GaAs devices do not suffer a penalty in IQE with deep plasma etch processes. As GaAs has by nature higher surface recombination rate than InP [20] and thus seems likely to be more prone to plasma etch related damage, we make the assumption that the difference in IQEs for the InP devices in Figure 7 (a) is related to the observed surface corrosion roughness.

In conclusion, from the observations from the LIV measurement data and the analysis presented above we see that the Euler U-bend strip device function is well preserved, while the addition of the rib-strip interface causes perturbations in the device operation.

### 3.3 Spectra and far-field

The spectra and far field measurements, presented in Figure 8 and Figure 9 respectively, have been made at a 150 mA injection current. We can see that while with the pure rib or strip devices the spectra exhibit an even Fabry-Perot comb structure, the rib-strip U-bend device suffers from a significant perturbation in the spectrum. A simulation using the finite-difference time-domain method estimated only a 0.22% reflection for the fundamental TE-mode at the rib-strip interface, implying that the resulting perturbation are related to an additional mechanism. The suspected reason for this is the slight presence of higher order modes (HOM) in the less strictly guided rib waveguide, further enhanced by the presence of the rib-strip interface. We can see the presence of HOMs in the far-field measurements in Figure 9. The strip waveguides, both straight and U-bend (Figure 9 (b) and (e)), exhibit single mode operation only; the equal width of the fast and slow axis can be contributed to the strong confinement of the lateral mode. This observation confirms that the Euler U-bend geometry itself can support the single-mode operation. The problem arises with the rib geometry where even the straight waveguide exhibits a faint trace of a secondary peak (Figure 9 (a)). The presence of the second peak becomes more evident with the rib-strip U-bend device (Figure 9 (c)), though Figure 9 (d) displays for a comparison a single peak plot that demonstrates that also single-mode only devices were present. This indicates that the waveguide width (2.9 µm) is borderline in its capacity to support the single-mode operation. From these observations it becomes evident that for a good quality U-bend device the rib-strip interface should be fabricated using tapering so that the mode can be adiabatically transferred from one waveguide geometry to other. Alternatively, by utilizing narrower waveguides, it is possible that the problem could be alleviated even with the simple straight rib-strip interface.

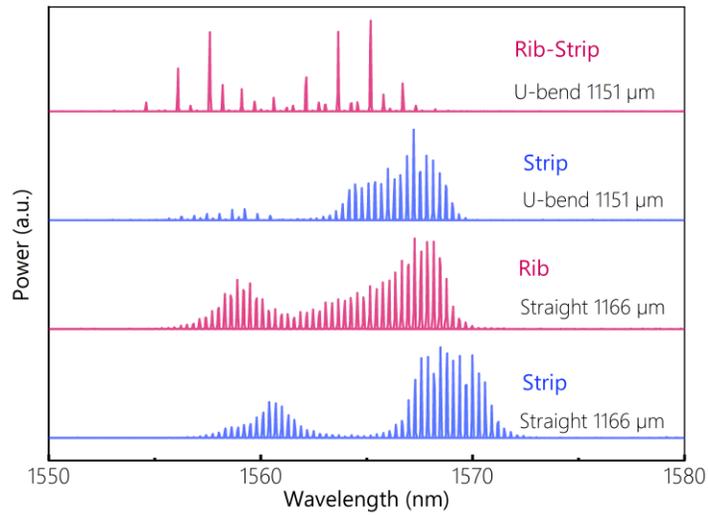

Figure 8 Spectra from all the LD device variants at 150 mA.

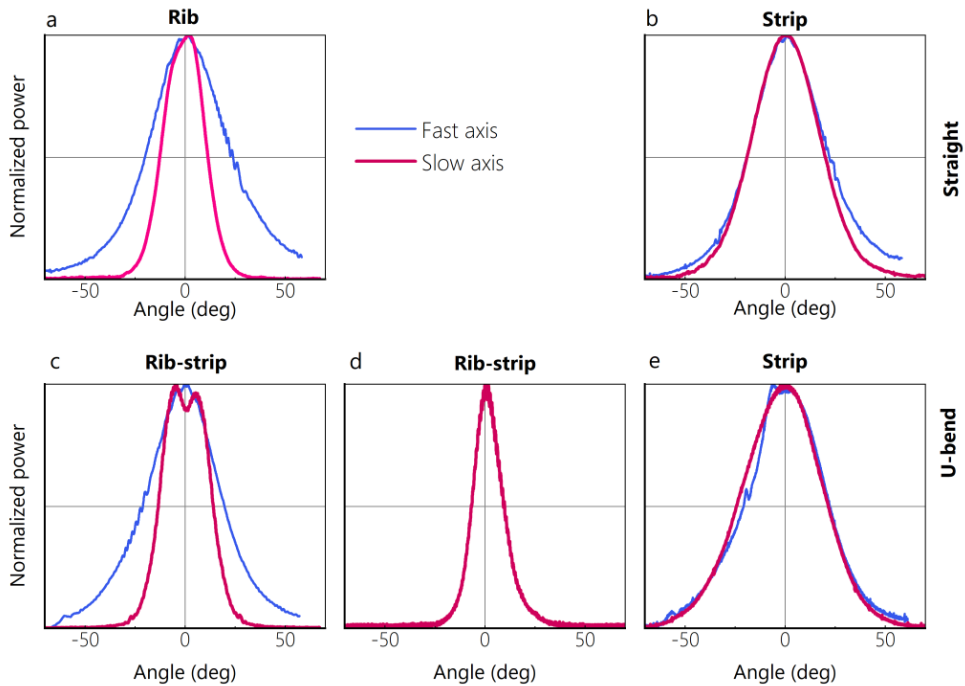

Figure 9 Far-field measurements from the LD device variants. a) Straight rib with a trace of a HOM, b) straight strip in single mode, c) U-bend rib-strip with HOMs d) U-bend rib-strip in single mode and e) U-bend strip in single mode.

### 3.4 SOA and flip-chip integration

The SOA U-bend devices were tested first without flip-chip integration by measuring the amplified spontaneous emission (ASE) and the IV data in addition to the spectral measurement similarly to the LD test devices. Figure 10 displays the measurement results for the 1150 µm cavity length devices with both strip and rib-strip geometries. From the ASE output power measured, we can see that while output power levels are similar, the rib-strip device is more prone to saturation and nonlinear behavior, as expected. The IV data shows no significant difference in performance. The spectrum for the strip waveguide device shows an expected wider broadband spectrum. The rib-strip spectrum on the other hand displays some additional lasing peaks, indicating that the interface forms a secondary cavity inside the device, as also suggested by the LD device characterization.

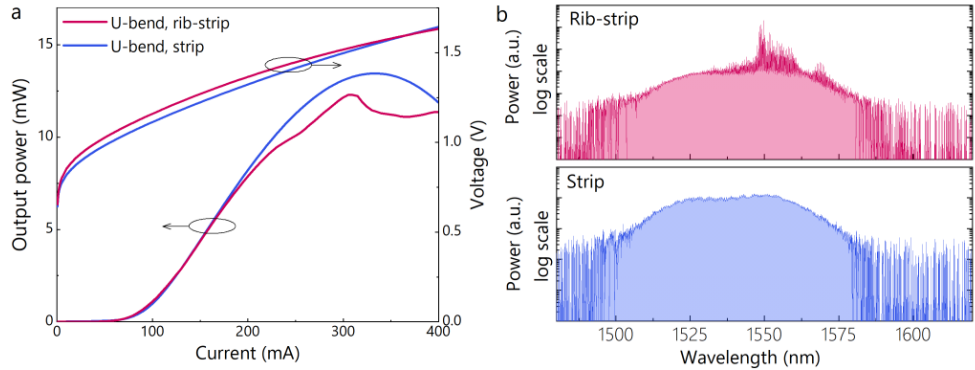

Figure 10 a) ASE power measurement and IV-curve from a U-bend SOAs before flip-chip integration. b) SOA spectra at 150 mA.

Finally, the U-bend SOA devices were flip-chip bonded on a 3 µm core SOI platform with waveguides for guiding the signal from the SOA to the opposing sides of the chip as illustrated in Figure 11. This way we can be sure that the signal measured from the front facet contains only the signal coupled into the waveguide and not any traces of uncoupled light. The devices were measured by manually probing the contacts with prober needles and light collected with an integrating sphere for the LI measurement and with a multimode fiber for the spectral measurement.

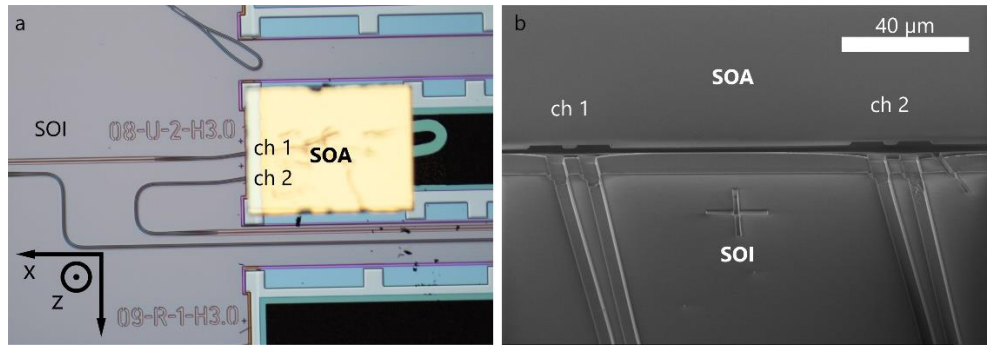

Figure 11 a) Flip-chip integrated U-bend SOA on SOI platform. b) SEM image of the SOA waveguide alignment with the SOI waveguides.

Figure 12 displays the measurement results from an integrated rib-strip U-bend SOA chip from both front and back facets. Out of the batch of the integrated test devices with both strip and rib-strip waveguides this chip had the best alignment and coupling efficiency. The ASE power data shows a decent ratio between the front and back facet signals though it seems that there is still some light escaping the coupling interface, based on the larger signal from front facet. In general, it was found that the III-V chip z-alignment (coordinate system noted in Figure

11 (a)) control was challenging in this trial and the integration yield of properly coupled devices was low. This can be attributed to debris and dicing damage on the integration interface. To solve this problem, the dicing quality and interface design can be improved. The spectra from both facets are similar with each other, displaying the wide ASE spectra and similarly with the unintegrated rib-strip device display a lasing peak that can be attributed to the rib-strip interface. The flip-chip integrated device exhibits an ASE power at 150 mA of a magnitude which is over 3 times the power of an unintegrated device at the same current. This demonstrates that the device withstands the integration process and that the p-side down integration significantly improves the thermal behavior leading to a higher output power.

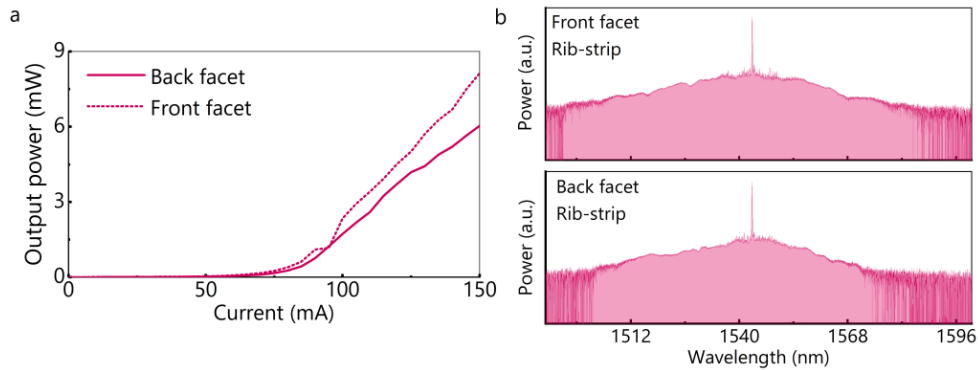

Figure 12 a) ASE power measurement and IV-curve from a rib-strip U-bend SOA after flip-chip integration measured from SOI front and back facet. b) SOA spectra at 150 mA from the front and back facets.

## 4 Conclusions

Euler U-bend TW gain devices based on InP material system were demonstrated. The focus was on a systematic analysis of the interplay between device functionality and the specific process steps used to optimize the operation of the U-bend gain waveguides. We estimate a bend loss of 0.56 dB/180° for a 2.9 µm wide waveguide with an effective bending radius of 50 µm at 1550 nm wavelength, while maintaining single mode operation in a strip waveguide geometry. Our analysis reveals that while the deep etched devices maintain an acceptable level of performance, to achieve the best possible performance a rib-strip device with a tapered interface should be utilized in the future work. Without tapering, a straight rib-strip interface causes a significant reduction of device output power and perturbations in the spectral and far field operation. Finally, we demonstrate the Euler U-bend structure in SOA flip-chip integrated on a 3 µm SOI platform. The performance of the InP/SOI gain structure is superior to the unintegrated devices owing to the deployment of p-side flip-chip bonding that exhibits improved thermal behavior.

**Conflict of interest**

The authors declare no conflicts of interest.

**Data availability statement**

The data that supports the findings of this study is available from the corresponding author upon reasonable request.